\documentclass{elsart1p}
\usepackage{graphicx}
\usepackage{amssymb}
\begin{document}
\begin{frontmatter}
%
%
%
%
%
\title{Heavy flavour suppression and flow}
%
%

\author{Santosh K Das and Jan-e Alam }

\address{Variable Energy Cyclotron Centre, 1/AF, Bidhan Nagar, Kolkata 700064, India}

\begin{abstract}
The drag and diffusion coefficients of charm and bottom quarks propagating
through quark gluon plasma have been evaluated for conditions
relevant to  nuclear collisions at Large Hadron Collider 
and Relativistic Heavy Ion Collider.
Both the radiative and collisional processes of energy loss are included
in evaluating the {\it effective} drag and diffusion coefficients. 
The  Landau-Pomeronchuk-Migdal (LPM) and the dead cone effects on the
radiative energy loss of heavy quarks have been included. 
With the effective transport coefficients the Fokker Plank (FP)
equation has been solved for the evolution of heavy quarks in QGP.
The solution of the FP equation has been used to
evaluate the nuclear suppression factor, $R_{\mathrm AA}$  
and the elliptic flow, $v_2^{HF}$
for the non-photonic single electron spectra resulting from the 
semi-leptonic decays 
of hadrons containing charm and bottom quarks. It is observed that
the experimental data from RHIC on $R_{\mathrm AA}$  and $v_2^{HF} $ 
can be reproduced simultaneously within the 
pQCD framework for the same set of inputs. 
\end{abstract}

\end{frontmatter}

\section{Introduction}
\label{}
The nuclear collisions at Relativistic Heavy Ion Collider (RHIC) 
and the Large Hadron Collider(LHC) energies 
are aimed at creating a phase where the bulk properties of the matter 
are governed by the (light) quarks and gluons. Such a 
phase of matter is called Quark Gluon Plasma (QGP). 
The study of the bulk properties of QGP is a field of great contemporary
interest and the heavy flavors, mainly, charm and bottom quarks, play a 
crucial role in such studies. As the relaxation time for heavy quarks 
are larger than the corresponding quantities
for light partons,
the light quarks and the gluons get thermalized 
faster than the heavy 
quarks. Therefore, the  propagation of  heavy quarks through 
QGP may be treated as the 
interactions between equilibrium and non-equilibrium degrees
of freedom and the Fokker-Planck (FP) equation provides an appropriate
framework~\cite{theory} for such studies. In this work we would like
to evaluate the elliptic flow and nuclear suppression factor of 
heavy flavours within the framework of FP equation. The results of
the calculations will be compared with the available experimental data.

The evolution of heavy quarks momentum distribution function, while propagating
through the QGP are assumed to be governed by the FP 
equation, which reads,

\begin{eqnarray}
\frac{\partial f}{\partial t} = 
\frac{\partial}{\partial p_i} \left[ A_i(p)f + 
\frac{\partial}{\partial p_j} \lbrack B_{ij}(p) f \rbrack\right] 
\label{expeq}
\end{eqnarray}
where the kernels $A_i$ and $B_{ij}$ are given by
\begin{eqnarray}
&& A_i = \int d^3 k \omega (p,k) k_i \nonumber\\
&&B_{ij} = \int d^3 k \omega (p,k) k_ik_j.
\end{eqnarray}
for $\mid\bf{p}\mid\rightarrow 0$,  $A_i\rightarrow \gamma p_i$
and $B_{ij}\rightarrow D\delta_{ij}$, where $\gamma$ and $D$ stand for
drag and diffusion co-efficients respectively.

The basic inputs required for solving the FP equation
are the dissipation co-efficients and initial momentum 
distributions of the heavy quarks. The drag and diffusion
coefficients have been evaluated by taking in to account
both the collisional and radiative processes~\cite{Das3}.
In the radiative process the dead cone
and LPM effects are included~\cite{Das3}.
In evaluating the drag co-efficient we have
used temperature dependent  strong coupling,
$\alpha_s(T)$~\cite{Kaczmarek}.
The Debye mass, $\sim g(T)T$ also a temperature dependent
quantity used as  a cut-off to shield the infrared divergences
arising due to the exchange of massless gluons.
The initial momentum distribution of heavy quarks 
in pp collisions have been taken from the
NLO MNR~\cite{MNR} code. The solution of the FP equation for the heavy  
quarks is convoluted with the fragmentation functions of the 
heavy quarks to obtain the transverse momentum ($p_T$)
 distribution of the 
$D$ and $B$ mesons. For heavy-quark fragmentation function, the
Peterson function has been used. The $p_T$ distribution 
of the electrons from the semi-leptonic decays of $D$ and $B$ meson
are evaluated using the standard techniques available in 
the literature. The ratio of the $p_T$ distribution of the 
electron from the decays of heavy flavours  produced  in heavy ion collisions 
to the corresponding (appropriately scaled by the number of collisions) 
quantities from the pp collisions is defined as the nuclear suppression 
factor, $R_{\mathrm AA}$:
\begin{eqnarray}
R_{AA}(p_T)=\frac{\frac{dN^e}{d^2p_Tdy}^{\mathrm Au+Au}}
{N_{\mathrm coll}\times\frac{dN^e}{d^2p_Tdy}^{\mathrm p+p}}
\label{raa}
\end{eqnarray}
which will be unity in the absence of re-scattering.
The STAR~\cite{stare} and the PHENIX~\cite{phenixe} collaborations have
measured the $R_{\mathrm AA}$ for non-photonic single electron as a function of
$p_T$ for Au+Au at $\sqrt{s_{NN}}=200$ GeV.

\begin{figure}
\begin{center}
\includegraphics[scale=0.35]{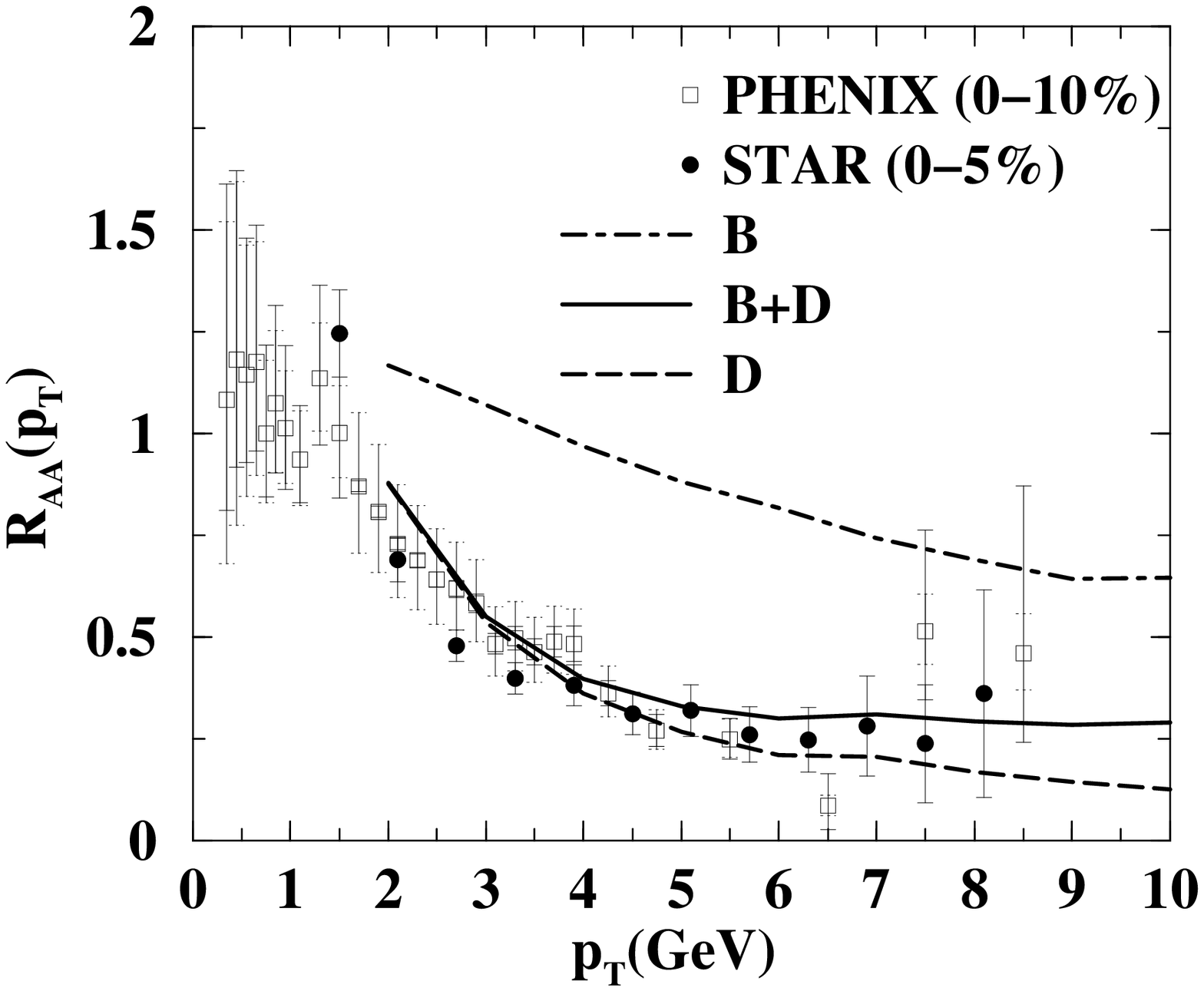}
\includegraphics[scale=0.35]{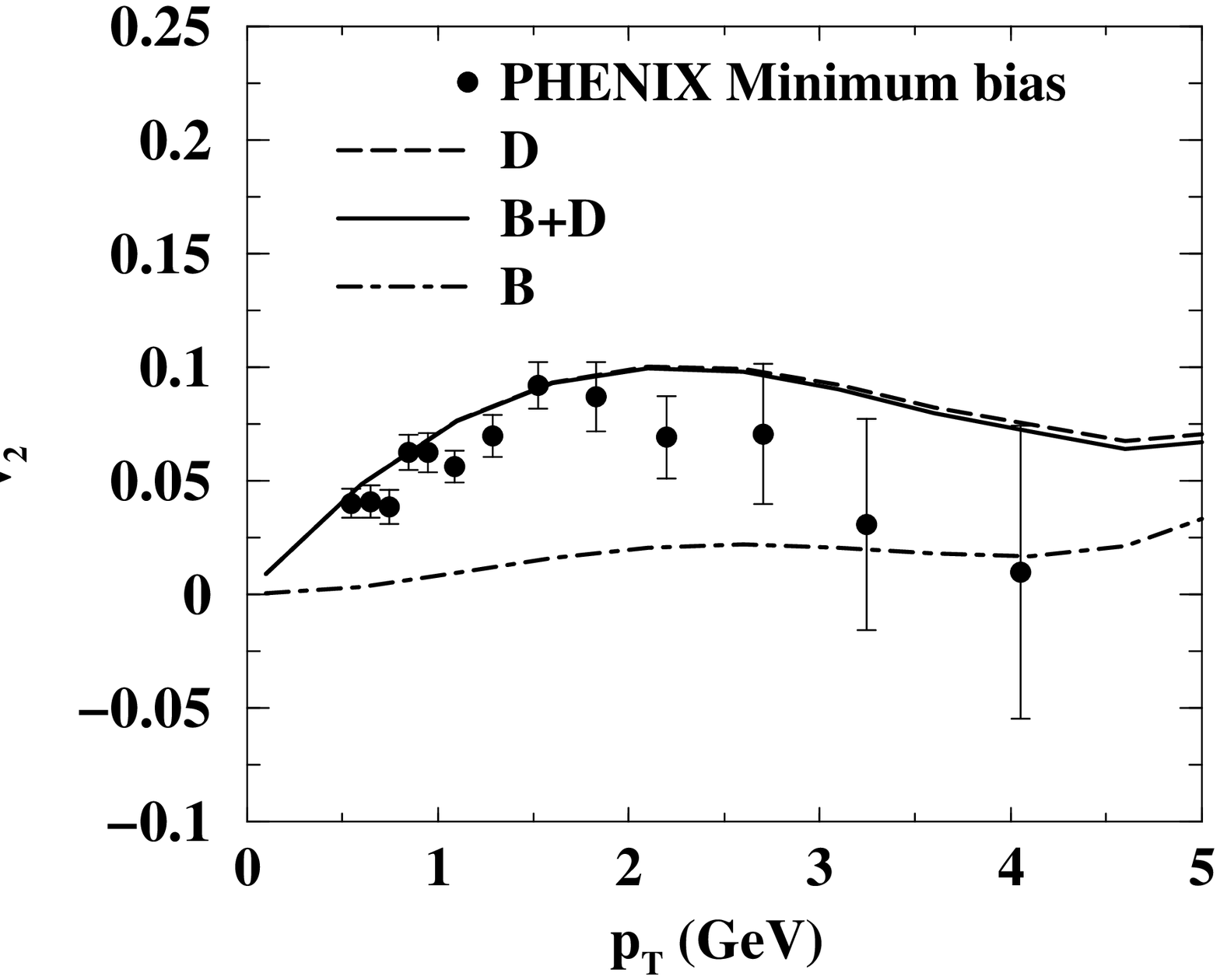}
\caption{The variation of $R_{\mathrm AA}$ and $v_2^{HF}$ with $p_T$. The 
theoretical results are obtained within the framework of FP equation described
in the text with the same set of inputs. The drag and diffusion coefficients
have been calculated using pQCD approach. 
Contributions from charm and bottom quarks are shown separately.
The initial temperature and thermalization time are
taken as 400 MeV and 0.2 fm/c respectively.
}
\label{fig1}
\end{center}
\end{figure}
The experimental
data from both the collaborations~\cite{stare,phenixe}
show $R_{AA}<1$ for $p_T\geq 2$ GeV indicating
substantial  interaction of the heavy quarks with
the plasma particles. The spectra evaluated using the formalism
described above reproduces the data reasonably well (Fig.~\ref{fig1}).
It is clear from the results displayed in Fig.~\ref{fig1} that
the charm quarks suffer more suppression than bottom quarks at
RHIC energy.

Now we discuss the elliptic flow  resulting from non-central
collisions of nuclei.
When a heavy quark propagates along the major
axis of an ellipsoidal domain of QGP (resulting from the non-central collisions)
then the number of interactions it encounters or in other words
the amount of energy it dissipates or the amount of momentum degradation
that takes place is different from when it propagates
along the minor axis. 
Therefore, the momentum distribution of electrons
originating from the decays of heavy flavoured hadrons produced
from the fragmentation of heavy quarks  propagating through
an anisotropic domain will reflect such anisotropy. 
Thus the spatial anisotropy due to non-central collisions
will be reflected through  the momentum distributions.  The degree
of momentum anisotropy will depend on both the spatial anisotropy and
more importantly on the coupling strength of the interactions 
between the heavy quarks and the QGP.  
The drag and diffusion coefficients depend on the temperature of
the background medium (QGP) which evolves in space and time due 
to expansion. Therefore, the drag and diffusion  will also change
due to the flow of the background.
The flow of the background has been treated
within the ambit of (2+1) dimensional hydrodynamics~\cite{hydro}.
The coefficient of elliptic 
flow, $v_2^{HF}$ is defined as:
\begin{eqnarray}
v_2^{HF}(p_T)=\langle cos(2\phi) \rangle= \frac{\int d\phi \frac{dN}{dydp_Td\phi×}|_{y=0} cos(2\phi)}
{\int d\phi \frac{dN}{dydp_Td\phi×}|_{y=0}×}
\end{eqnarray}
We evaluate $v_2^{HF}$ in the current formalism~\cite{Das4} (see also
~\cite{rappv2}) and compare 
the results with experimental data~\cite{phenixemb}
(Fig.~\ref{fig1},  right panel). 
\begin{figure}
\begin{center}
\includegraphics[scale=0.35]{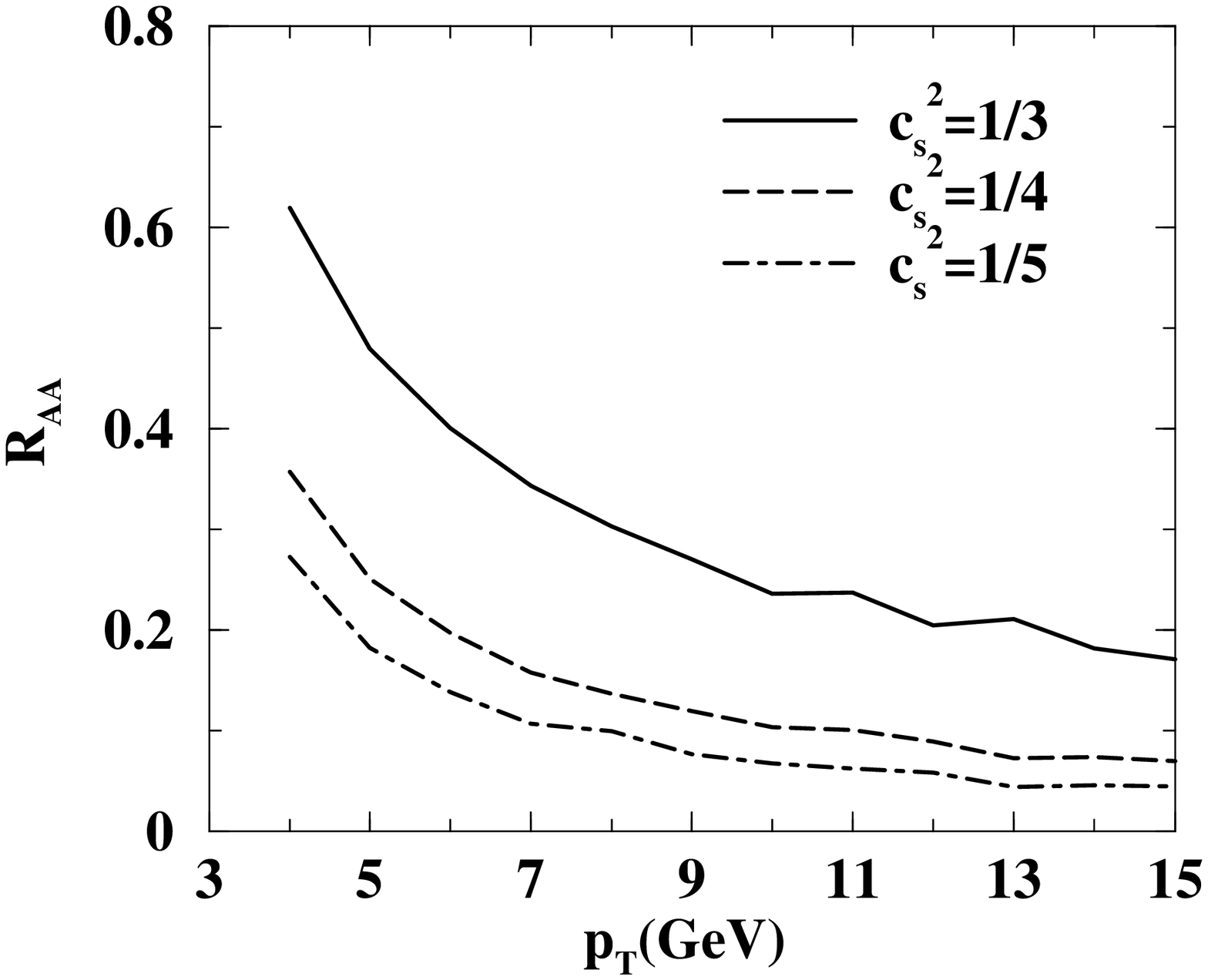}
\includegraphics[scale=0.35]{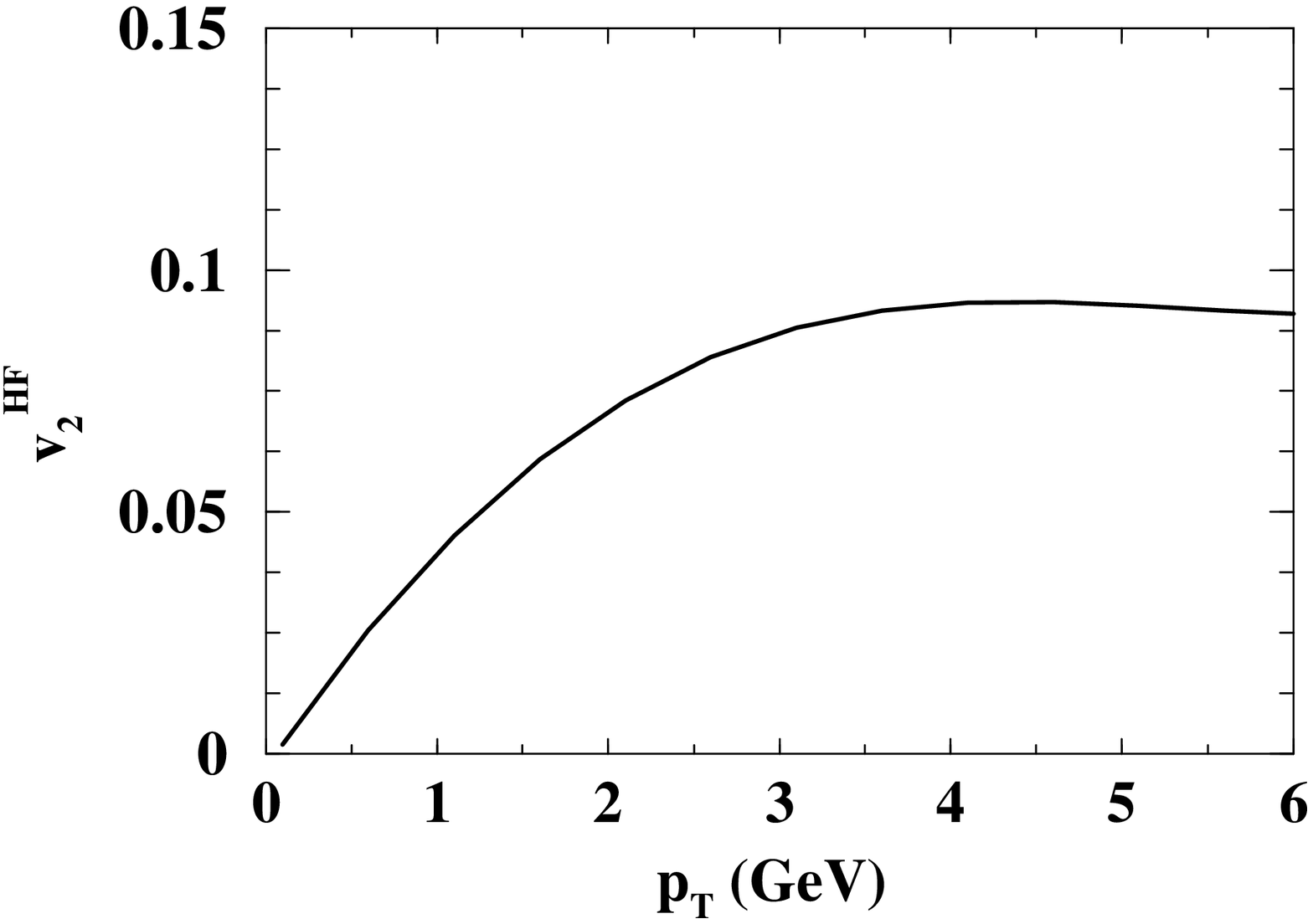}
\caption{The variation of $R_{\mathrm AA}$ and $v_2^{HF}$ with $p_T$ for 
LHC for 0-10\% centrality. The initial temperature and thermalization time are
taken as 700 MeV and 0.08 fm/c respectively.
The theoretical results are obtained within the framework of FP equation 
described
in the text with the same set of inputs. The drag and diffusion coefficients
have been calculated using pQCD approach.}
\label{fig2}
\end{center}
\end{figure}

The prediction for the nuclear suppression factor and elliptic
flow of the heavy quarks to be measured at LHC energy 
through the semi-leptonic decays are depicted in Fig.~\ref{fig2}. 
The sensitivity of the results on the equation of state (EoS),
i.e. on the velocity of sound is also considered.
Lowering of $c_s$ gives more suppressions as observed in Fig.~\ref{fig2}. 
Lower value of velocity of sound,  $c_s$ makes the expansion of the plasma 
slower enabling the propagating heavy quarks to spend more time to interact
in the medium and hence lose more energy before exiting from the plasma
which results in less particle production at high $p_T$. 

Some comments on the $R_{\mathrm AA}$  vis-a-vis $v_2^{HF}$
are in order here.  The 
$R_{\mathrm AA}$  contains the ratio of $p_T$ distribution 
of electron resulting from Au+Au to p+p collisions, where
the numerator contains the interaction of the heavy quarks
with the medium (QGP) and such interactions are absence in 
the denominator. Whereas for $v_2^{HF}$ both the numerator and the
denominator contain the interactions with the medium, resulting in
some sort of cancellation. Therefore, the $K$ factor dependence 
for the two cases seems to be different.

In summary, we have evaluated the nuclear suppression,
$R_{AA}$ and the azimuthal asymmetries
($v_2^{HF}$) using the Fokker-Planck  equation.
The transport coefficients are evaluated using pQCD interaction.
We have shown that our results on both the $R_{\mathrm AA}$ and $v_2^{HF}$ 
can reproduce the experimental data with the same set of inputs.

{\bf Acknowledgment:}
This work is supported by DAE-BRNS project Sanction No. 2005/21/5-BRNS/2455.


\begin{thebibliography}{00}
\bibitem{theory} B. Svetitsky, Phys. Rev. D {\bf 37}, 2484( 1988);
 J. Alam, S. Raha and B. Sinha, Phys. Rev. Lett. {\bf 73}, 1895 (1994);
 G. D. Moore and D. Teaney, Phys. Rev. C {\bf 71}, 064904 (2005);
 H. van Hees, R. Rapp, Phys. Rev. C,{\bf 71}, 034907 (2005);
S. K Das, J. Alam and P. Mohanty, Phys. Rev. C {\bf 80}, 054916 (2009);
S. K Das, J. Alam, P. Mohanty and B. Sinha
Phys. Rev. C {\bf 81}, 044912 (2010);
S. K Das, J. Alam and P. Mohanty, arXiv:0912.4089 [nucl-th]. 

\bibitem{Das3}S. K Das, J. Alam and P. Mohanty, 
Phys. Rev. C {\bf 82}, 014908 (2010).


\bibitem{Kaczmarek} O. Kaczmarek and F. Zantow, 
Phys. Rev. D, {\bf 71}, 114510 (2005).

\bibitem{MNR} M. L. Mangano, P. Nason and G. Ridolfi, Nucl. Phys.
B {\bf 538}, 282 (2002).

\bibitem{stare} B. I. Abeleb {\it et al.} (STAR Collaboration), Phys. Rev. 
Lett. {\bf 98}, 192301 (2007).

\bibitem{phenixe} S. S. Adler {\it et al.} (PHENIX Collaboration), 
Phys. Rev. Lett. {\bf 96}, 032301 (2006).

\bibitem{hydro} P. F. Kolb, J. Sollfrank and U. Heinz, 
Phys. Rev. C {\bf 62}, 054909 (2000); P. F. Kolb and R. Rapp, Phys. Rev. 
C {\bf 67}, 044903 (2003); J. Sollfrank, P. Koch and U. Heinz,
 Phys. Lett. B {\bf 252}, 256 (1990); J. Sollfrank, P. Koch and 
U. Heinz, Z. Phys. C {\bf 52}, 593 (1991).

\bibitem{Das4}S. K Das and J. Alam, arXiv:1008.2643 [nucl-th].

\bibitem{rappv2} H. van Hees, V. Greco and R. Rapp, Phys. Rev. C {\bf 73},
 034913 (2006).

\bibitem{phenixemb} S. S. Adler {\it et al.} (PHENIX Collaboration),
Phys. Rev. Lett. {\bf 98}, 172301 (2007).

\end{thebibliography}
\end{document}